\documentclass[a4paper,11pt]{article}
\usepackage{jinstpub} 
\usepackage{lineno}

\title{Development of a picosecond-timing Cherenkov detector using gaseous photomultiplification}

\author[a]{R. Okubo}
\author[abc]{, K. Matsuoka}
\author[a]{, T. Iijima}
\author[ab]{, K. Inami}
\author[a]{, Y. Horii}
\author[a]{, K. Suzuki}
\author[a]{, K. Sumi}
\author[a]{, T. Ichikawa}
\author[a]{, K. Ueda}
\author[a]{, S. Koji}
\author[a]{, A. Kondo}
\author[a]{, and K. Chiba}
\affiliation[a]{Nagoya University, Nagoya, Japan}
\affiliation[b]{High Energy Accelerator Research Organization (KEK), Tsukuba, Japan}
\affiliation[c]{The Graduate University for Advanced Studies (SOKENDAI), Hayama, Japan}

\emailAdd{rokubo@hepl.phys.nagoya-u.ac.jp}

\abstract{
Photosensitive gaseous detectors with a simple photoelectron multiplication mechanism as resistive plate chambers are expected to offer both large photo coverage and excellent time resolution while keeping costs low. We have developed a gaseous photomultiplier (GasPM) and demonstrated that a single-photon time resolution is $25\pm1.1~\rm{ps}$ at the gain of $3.3\times10^6$ with a $\rm {LaB_6}$ photocathode, which has an extremely low quantum efficiency. We then developed a Cherenkov detector using GasPM with a $\rm{CsI}$ photocathode aiming for an application in time-of-flight measurements with a resolution below 10~$\rm{ps}$ for particle identification. We performed a test using the 3 GeV electron beam at the PF-AR test beamline located at KEK, Japan. As a result, the resolution of time-of-flight between the detector and an MCP-PMT is measured to be $\sigma=73.0\pm2.4~\rm{ps}$. The obtained resolution is worse than the target because of the lower gap voltage compared to the design. However, it is consistent with the expectation from the applied gap voltage. It is an important milestone for achieving the designed resolution by increasing the gap voltage and photon detection efficiency in future development.

}

\begin{document}
\maketitle
\flushbottom

\section{Introduction}
\label{sec:intro}
Scaling up the size and performance of detectors is inevitable for future particle and nuclear physics experiments to exploit the physics frontier. In particular, picosecond-timing photodetectors with a large photo coverage are required for upgrading Cherenkov detectors. For example, the time-of-propagation (TOP) counter~\cite{top} is a novel particle identification (PID) detector for the barrel region at the Belle II experiment. It identifies particles by measuring Cherenkov photon timing with 30~$\rm{ps}$ single photon time resolution using micro-channel-plate (MCP)-PMTs~\cite{top_mcp_pmt}. MCP-PMTs are also being considered for use in the ring-imaging Cherenkov detector upgrade at the LHCb experiment\cite{lappd_rich_upgrade}, where large area coverage with high time resolution is required. Large MCP-PMTs called large area picosecond photodetectors (LAPPDs) could satisfy such a requirement. The largest LAPPD has a $\rm{ 192\times192~mm^2 }$ sensitive area while keeping $\sigma=39~\rm{ps}$ single photon time resolution~\cite{lappd}. However, the high cost of MCP-PMTs will be a bottleneck for large-area coverage in the future.

To overcome this bottleneck, we have developed a resistive plate chamber (RPC) based gas detector named gaseous photomultiplier (GasPM) that is expected to have a large photo coverage, lower cost, and higher time resolution.
\section{Single photon response of GasPM}
\subsection{Detector design}
\label{sec:design_first_proto}
\begin{figure}[htbp]
    \centering
    \includegraphics[width=1.0\linewidth]{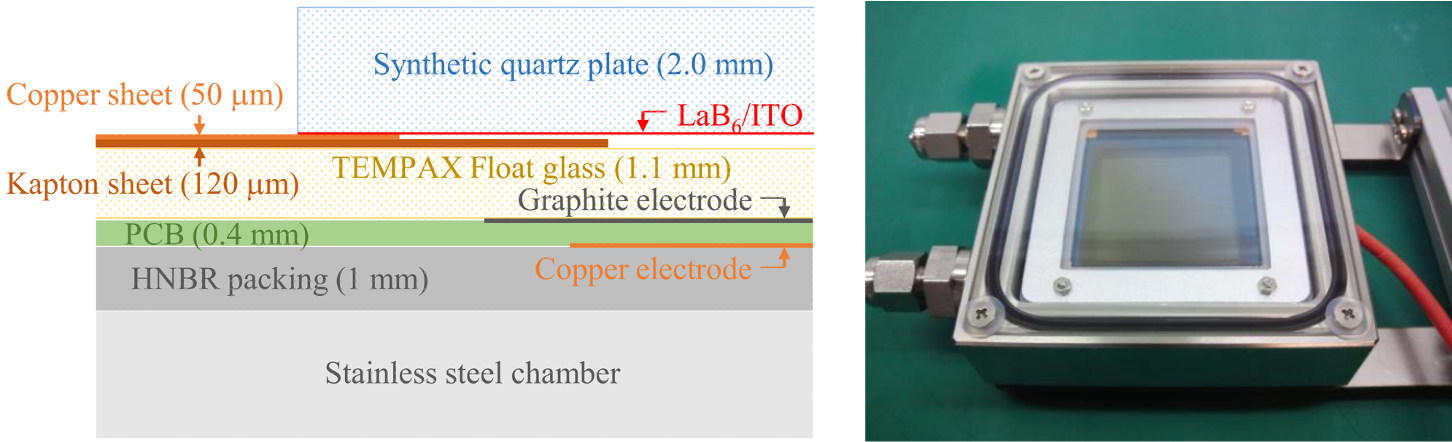}
    \caption{Design of the GasPM prototype (left)~\cite{gaspm} and a photo of the prototype (right)~\cite{gaspm}.}
    \label{fig:gaspm_1st}
\end{figure}

We first design a prototype of GasPM to demonstrate high time resolution. Figure~\ref{fig:gaspm_1st} shows the design and a photo of the prototype. Photoelectrons from the photocathode are rapidly amplified by avalanche multiplication in the narrow gas gap, and the amplified electrons induce current at the copper electrode. We apply a high voltage of $3.0~\rm{kV}$ to the narrow gas gap of 170~$\rm{\mu m}$ width. The gas gap is filled with a mixture of $90\%$ $\rm{C_2 H_2F_4}$ and $10\%$ $\rm{SF}_6$ gases. We use $\rm{LaB_6}$ for the photocathode, which has an extremely low quantum efficiency but stable performance in gases. The $\rm{TEMPAX~Float^{\text{\textregistered}}}$ resistive plate has a high resistivity of $10^{15}~\Omega\rm{cm}$ so it can prevent discharges in the gas gap.
\subsection{Detector response to single photons}
\label{sec:detector_response_to_single_photons}
\begin{figure}[htbp]
    \centering
    \includegraphics[width=1.0\linewidth]{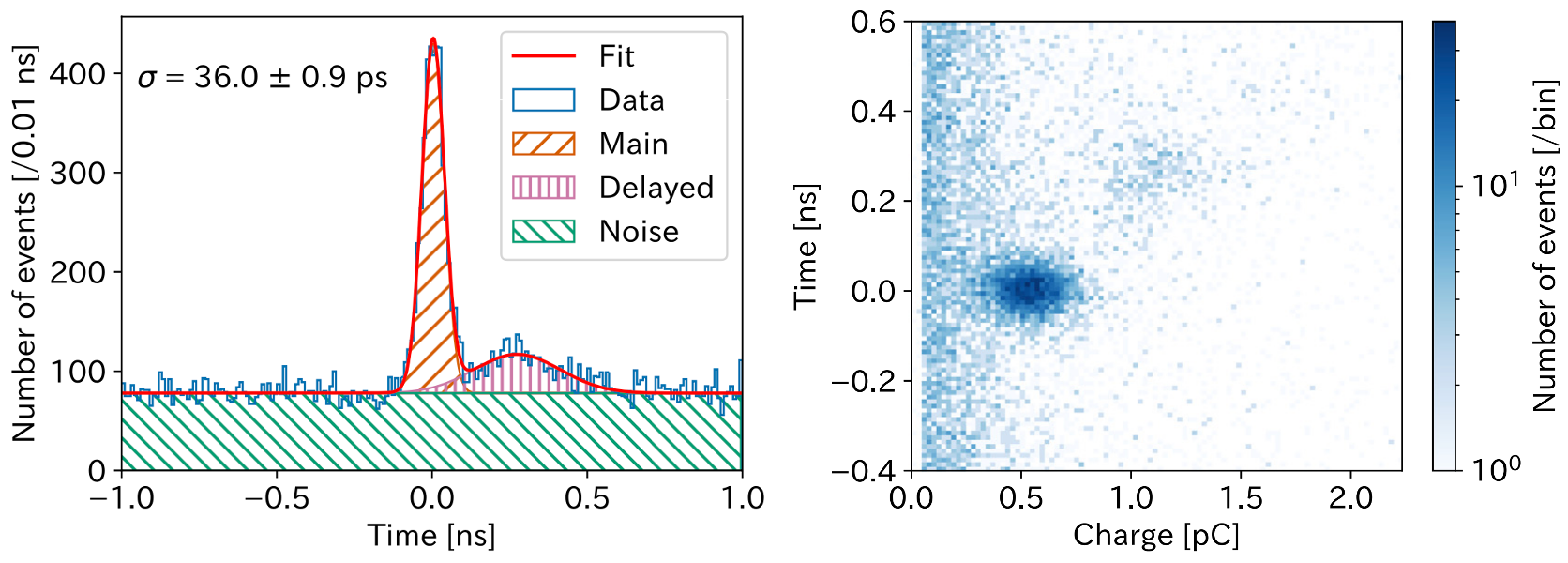}
    \caption{Timing distribution of the detector signals and fit projections (left)~\cite{gaspm}, and the time and charge distribution (right)~\cite{gaspm}. The time is calculated from the timing difference between the detector and the laser, and the time walk effect is corrected.}
    \label{fig:gaspm_laser}
\end{figure}

We measure the single photon time resolution of the prototype GasPM. We use a picosecond pulse laser with a wavelength of 375~nm and a width of $\sigma=22.3\pm0.5~\rm{ps}$. The time width of the laser is measured by the manufacturer using a streak camera. We irradiate the detector with the laser to measure the waveforms of the detector and laser synchronization signals using a DRS4 evaluation board, which is a digitizer with a sampling rate of 5~Gsamples/s. 
First, we measure the time resolution of the detector using the time difference between the detector and laser synchronization signals. The timing of the detector signal is extracted by fitting a polynomial function to the rising edge of the waveform. We extract the pedestal-corrected pulse height from the fit and then extract the timing when the waveform exceeds 50\% of the pulse height. The laser timing is also estimated in the same way. Figure~\ref{fig:gaspm_laser} shows the timing and the charge distributions of the detector. We observe a peak of main signals, and the delayed signals with a larger gain are also observed. The random timing signals are considered to be random noise or signals due to cosmic rays. The delayed signals are shifted by about $200$-$300~\rm{ps}$ from the main peak. We consider that the delayed signal consists of two or more avalanches due to feedback of photons from the original avalanche. The charge of the delayed signals is typically twice that of the main signals, and the time delay is equal to the drift time in the gas gap, which supports our view. We estimate the photon feedback occurrence to be $0.30\pm0.02$ by fitting the time and charge distributions assuming a Poisson model. We estimate the time resolution of GasPM using only the main signals because the degradation of the resolution by the delayed signals can be improved in the future by suppressing them or using a digitizer with a higher sampling rate. We fit the time distribution with a sum of double Gaussian functions for the main and the delayed signals and a constant function for the random timing noise. The time resolution of the detector is estimated to be $\rm{25.0\pm1.1~ps}$~\cite{gaspm} after subtracting $\sigma=22.3\pm0.5~\rm{ps}$ of the laser width and $\sigma=14.0\pm0.3~\rm{ps}$ of the readout resolution from $\sigma= 36.0\pm 0.9~\rm{ps}$ of the main peak width.

\section{Development of a Cherenkov timing detector}
\subsection{Detector design}
We develop a prototype of a Cherenkov timing detector utilizing GasPM. We change the photocathode from $\rm{LaB_6}$ to $\rm{CsI}$, which is known to have a sufficient quantum efficiency for UV light and a high resistance to the gases used in GasPM. We also change the window of GasPM from quartz to 2.4~mm thick $\rm{MgF_2}$ for a better transmission of ultraviolet (UV) light to which the CsI photocathode is sensitive. The thickness of the gas gap is set to $200~\rm{\mu m}$. We estimate that 8 photons per track could be detected with this configuration, and the time resolution could reach about $\rm{9~ps}$, estimated from $\sigma=25.0~{\rm ps}/\sqrt{N}$, where $N$ is the number of photons detected per track. Such a detector is useful as a picosecond timing time-of-flight (TOF) detector for PID in particle physics experiments. For example, if this detector is used to precisely measure the TOF at the Belle II experiment, the PID performance is expected to be improved to $99.7\%$ kaon efficiency with $0.4\%$ pion misidentification probability for 2~$\rm{GeV/c}$ kaons. The detector is also sensitive to low-energy gamma rays by detecting the electrons produced by Compton scattering in the $\rm{MgF_2}$ window, and the gamma ray timing could help to reduce false electromagnetic calorimeter clusters due to beam-induced background.

\subsection{Detector response to charged particles}
\begin{figure}[htbp]
\begin{minipage}[t]{0.32\columnwidth}
    \includegraphics[width=\columnwidth]{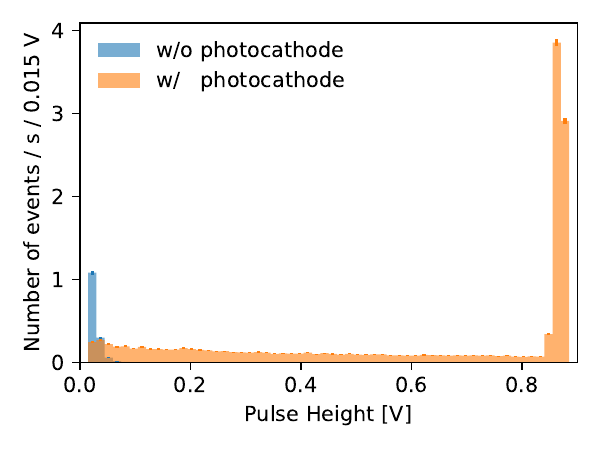}
    \caption{Distributions of the pulse height with and without $\rm{CsI}$ photocathode. 
    The peak around 0.9~V corresponds to signals exceeding the maximum input voltage range of the readout.}
    \label{fig:height_distrib}
\end{minipage}
\hspace{0.03\columnwidth} 
\begin{minipage}[t]{0.65\columnwidth}
    \centering
    \includegraphics[width=\columnwidth]{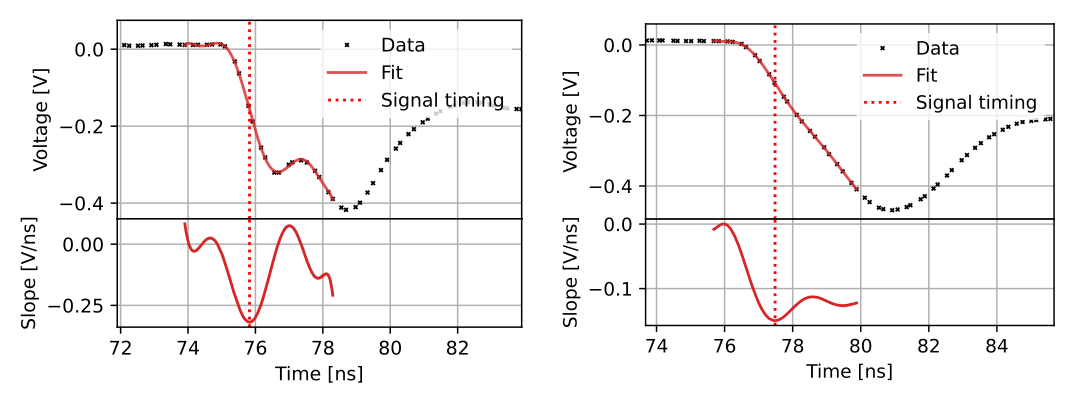}
    \caption{Waveforms from the GasPM. The upper panel shows the raw data and the rising edge fit, and the lower panel shows the slope of the fitted curve. On the left is a typical event without pulse overlap, and on the right is a typical pulse overlap event.}
    \label{fig:wfm_fit}
\end{minipage}
\end{figure}

We evaluate the performance of the prototype of the GasPM Cherenkov detector using a 3~GeV electron beam at the PF-AR test beamline located at KEK, Japan. We apply $\rm{2.8~kV}$ to the gas gap in this test. We trigger the beam by a coincidence of two scintillation counters. We first measure the signal detection rate of the GasPM to demonstrate Cherenkov photon detection by comparing the rate to an RPC of the same design as the GasPM with the photocathode replaced by an electrode. Figure~\ref{fig:height_distrib} shows the pulse height distributions of the GasPM and RPC. We observe 1.5~Hz of signals for the RPC without photocathode, and 13.8~Hz for the GasPM with photocathode. The increase of the detection rate as well as the pulse height is attributed to Cherenkov light detection. The detected number of photons per event in the test is approximately one for the GasPM although we estimated it to be seven, and we expect the time resolution with the single photon detection with this settings to be about 50-70~$\rm{ps}$. 

\begin{figure}[htbp]
    \centering
    \includegraphics[width=1.0\linewidth]{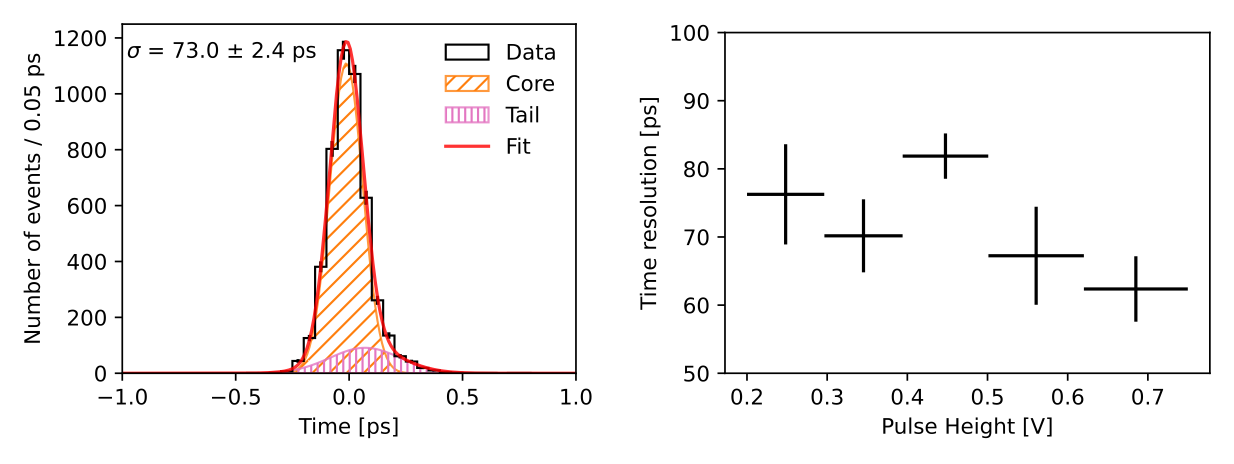}
    \caption{Time resolution of the TOF measured by the GasPM and MCP-PMT for the 3~GeV electron beam. The left plot shows the distribution of the TOF between the GasPM and MCP-PMT with the fit projection overlaid. The right plot shows the time resolution sliced by the pulse height of the GasPM. }
    \label{fig:resolution_beamtest}
\end{figure}

We measure the TOF between an MCP-PMT with a Cherenkov radiator and the GasPM. We also measure the TOF between two MCP-PMTs to estimate their time resolution, which is measured to be $15.8\pm1.1~\rm{ps}$ for each MCP-PMT. 
We extract the timing of the GasPM signal from the waveform. We take a different approach than in Sec.~\ref{sec:detector_response_to_single_photons} because we couldn't observe enough events without pulse overlap to evaluate the time resolution. We first extract the rising edge of the pulse and then fit it with a polynomial function. We extract the timing of the signal when the slope of the fitted curve first reaches the local maximum, ignoring the fluctuation of the baseline. The fit projections to the waveforms are shown in Figure~\ref{fig:wfm_fit}. This new approach can extract the timing of the first pulse from pulse overlap events without using the pulse height, which is difficult to be estimated for pulse overlap events.
Figure~\ref{fig:resolution_beamtest} shows the time resolution of the TOF measured by the GasPM and MCP-PMT. We fit the distribution by two Gaussian functions. The resolution of the core component is $73.0\pm2.4~\rm{ps}$ for all signals and $62.3\pm4.8~\rm{ps}$ at the largest bin of the pulse height. It is consistent with the expected time resolution with this setup

\subsection{Future prospects of the detector}
We plan to improve the detector for a better time resolution. The number of detected photons can be improved by a factor of ten by using a two-times-thicker Cherenkov radiator and a photocathode with a better quantum efficiency. We also plan to increase the gap voltage from 2.8~kV to 3.5~kV, which will improve the single photon time resolution by a factor of two. With these improvements, the time resolution is expected to be less than 10~$\rm{ps}$. We also plan to increase the sampling rate of the digitizer from 5 to 10~Gsamples/s to improve the discrimination of the first pulse from the overlapping second pulse by the photon feedback.

\section{Summary}
We develop the RPC-based gaseous photomultiplier with the goal of realizing a photodetector with a large sensitive area, low-cost, and high-time resolution. We first develop the prototype using the $\rm{LaB_6}$ photocathode, which has extremely low QE but stable performance in gases and the air, and perform a test using the picosecond pulse laser. As a result, we demonstrate $\sigma=25.0\pm1.1~\rm{ps}$ of the single photon time resolution. We also observe delayed timing signals with larger pulse heights, which would be feedback photons from the first avalanche amplification. We then develop the Cherenkov timing detector using the GasPM with the $\rm{CsI}$ photocathode and $\rm{MgF_2}$ Cherenkov radiator. We estimate the performance using the $3~\rm{GeV}$ electron beam. The time resolution for the electron beam TOF between the detector and an MCP-PMT is measured to be $\sigma=73.0\pm2.4~\rm{ps}$ for all signals, and $\sigma=62.3\pm4.8~\rm{ps}$ for the high gain signals. These results are consistent with the expectation.

\section{Acknowledgement}
This work was supported by DAIKO FOUNDATION and MEXT/JSPS KAKENHI Grant Numbers JP26610068, JP16H00865, JP19H05099, JP21H01091, and JP23H05433. We acknowledge the support of KEK in our test at the PF-AR test beamline.



\end{document}